\documentclass[twocolumn,showpacs,preprintnumbers,amsmath,amssymb]{revtex4}
\usepackage{graphicx}
\usepackage{dcolumn}
\usepackage{bm}

\begin{document}
\title{Higher order coherence of exciton-polariton condensates}
\author{Tomoyuki Horikiri$^1$$^2$}
\author{Paolo Schwendimann$^3$} 
\author{Antonio Quattropani$^3$} 
\author{Sven H$\ddot{\rm o}$fling$^4$}
\author{Alfred Forchel, $^4$}
\author{Yoshihisa Yamamoto$^1$$^2$}
\affiliation{
$^1$National Institute of Informatics, Hitotsubashi, Chiyoda-ku, Tokyo 
101-8430, Japan\\
$^2$E. L. Ginzton Laboratory, Stanford university, Stanford, 
California 94305-4088, USA  \\
$^3$Institute of Theoretical Physics, $\acute{E}$cole Polytechnique 
F$\acute{e}$d$\acute{e}$rale de Lausanne, CH 1015 Lausanne-EPFL, 
Switzerland \\
$^4$Technische Physik, Universit$\ddot{a}$t W$\ddot{u}$rzburg, Am 
Hubland, D-97074 W$\ddot{u}$rzburg, Germany
 }%
\date{\today}
\begin{abstract}
The second and third order coherence functions $g^{(n)}(0)\ (n=2 \ {\rm 
 and }\ 3)$ of an exciton-polariton 
 condensate is measured and compared to the theory. Contrary to an ideal photon laser, deviation from 
 unity in the second and third order coherence functions is observed, 
 thus showing a bunching effect, but not the characteristics of a 
 standard thermal state with $g^{(n)}(0)=n!$. 
 The increase of bunching with the order of the coherence function, 
 $g^{(3)}(0) > g^{(2)}(0)>1$,
 indicates that the polariton condensate is different from 
 coherent state, number state and thermal state. The experimental results are in 
 agreement with the theoretical model where polariton-polariton 
 and polariton-phonon interactions are responsible for the loss of 
 temporal coherence.
\end{abstract}

\maketitle

 In a seminal paper by R. J. Glauber marking the beginning of quantum 
 optics, coherent states were studied as unique states, whose normalized 
 coherence functions $g^{(n)}(0)$ are unity for all orders $n$ \cite{glauber}. 
 It has been shown theoretically and experimentally that a photon laser is well 
 characterized by coherence functions $g^{(n)}(0)$ close to one 
 at pump intensities well above threshold, thus showing all order 
 coherence, except for a random walk phase diffusion.
  Bose-Einstein condensation (BEC) of 
 polaritons has recently been the subject 
 of detailed investigations \cite{hui,kasprzak} and the question of 
 its coherence properties arises. BEC 
 is characterized by a large number of particles sharing the same 
 quantum state. Therefore it is expected that the state of polariton 
 condensates 
 shows temporal coherence properties leading to a coherence function of
 coherent state $g^{(n)}(0)=1$ or $N$-particle number state 
 $g^{(n)}(0)=1-{ n-1 \over N}$. However, there are important 
 differences between photon laser and polariton BEC, one of which 
 is that the former is based on 
 stimulated emission of photons from an excited level and population inversion is 
 necessary to reach a threshold, while the latter is
 based on stimulated cooling of polaritons to the ground state from 
 excited states and BEC threshold condition is given by $n\lambda_{T}^2\simeq 
 2\ln{L/\lambda_T}$, where $n$, $\lambda_T$ and $L$ are
 the polariton density, thermal de Broglie wavelength and system size, respectively. 
 Thus, when condensation occurs 
 in a lower polariton (LP) ground state, the population inversion is not 
 created yet in quantum wells \cite{imamoglu,sch}. Furthermore, 
 scattering processes to the non-condensed polariton modes may 
 influence  the coherence properties of the polariton condensate. This 
 last effect is absent in a photon laser. 
 Contrary to a photon laser, deviation from 
 unity in the second order coherence function $g^{(2)}(0)$ for the 
 exciton-polariton condensate has been observed \cite{hui,lausanne} 
 showing a bunching effect. The origin of the deviation from 
 unity has not been well identified so far. However, in 
 recent theoretical papers, the deviation 
 from full coherence in higher order correlations have been shown
 \cite{sch, sch2, haug}. The model presented in \cite{sch, sch2, haug}, considers 
 both polariton-polariton and polariton-phonon scattering. This 
 scattering processes lead to two competing effects that involve the 
 condensed polaritons: resonant scattering into and out of the polariton 
 ground state involving one polariton in the ground state and parametric 
 scattering of two polaritons in the ground state into two states of 
 opposite momentum. Below threshold, the statistics of the polaritons is 
 determined by the thermal phonon bath, which is responsible for 
 polariton relaxation, and the effect of polariton-polariton scattering 
 is negligible. In this case, the corresponding $g^{(n)}(0)$ is given 
 by that of a thermal state $n!$.  Above threshold, resonant polariton-polariton and 
 polariton-phonon interactions compete.
 The parametric 
 scattering   between ground state and excited states induces a 
 depletion of the ground state population. This effect induces 
 fluctuations in the population and manifests itself in a loss of 
 coherence in the polariton condensate \cite{sch, sch2}. Indeed, when 
 higher order coherence functions are calculated, a 
 deviation from full coherence and bunching effect above threshold are obtained.
 
 Let us discuss this point in more detail. The particle statistics of 
 a thermal state are characterized by the geometrical distribution 
 $p(m)={1\over \mu+1}({\mu \over \mu +1})^m$, where $p(m)$ is a probability of 
 $m$ polaritons and $\mu$ is mean polariton number, and lead to a 
 normalized coherence function $g^{(n)}(0)= n!$, which grows with $n$. In 
 this case, we expect a larger bunching behavior in the higher order 
 coherence functions, such as $g^{(3)}(0)>g^{(2)}(0)$. This behavior is indeed found for polaritons in the 
 ground state below threshold \cite{sch}. Above threshold, the 
 calculated particle distribution has a shape close to but different from 
 a Poissonian distribution.  The second order coherence 
 function is larger than one and the higher order coherence functions 
 are even larger \cite{sch}. 
 In this paper, we present the measurement results of the second and 
 third order normalized coherence 
 functions. In particular, we show that the observed behavior of growing 
 bunching with the order $n$ agrees well with the theory.

 The second order coherence function for exciton-polariton condensates has 
 been measured for 
 GaAs quantum well (QW)
 microcavity \cite{hui} and CdTe QW microcavity \cite{lausanne}. 
 Below condensation threshold, $g^{(2)}(0)$ takes almost unity since both
 polariton emission decay time below threshold and detector response 
 time are much longer than the intensity
 correlation time of polaritons.
 Above threshold $g^{(2)}(0)$ increases steeply 
 and then gradually saturates as the pump intensity goes up. 
 A full coherence $g^{(2)}(0)=1 $ has not been observed even at far 
 above threshold.

\begin{figure}[Hh]
   \begin{center}
    \rotatebox{0}{
    \scalebox{0.5}{
    \includegraphics{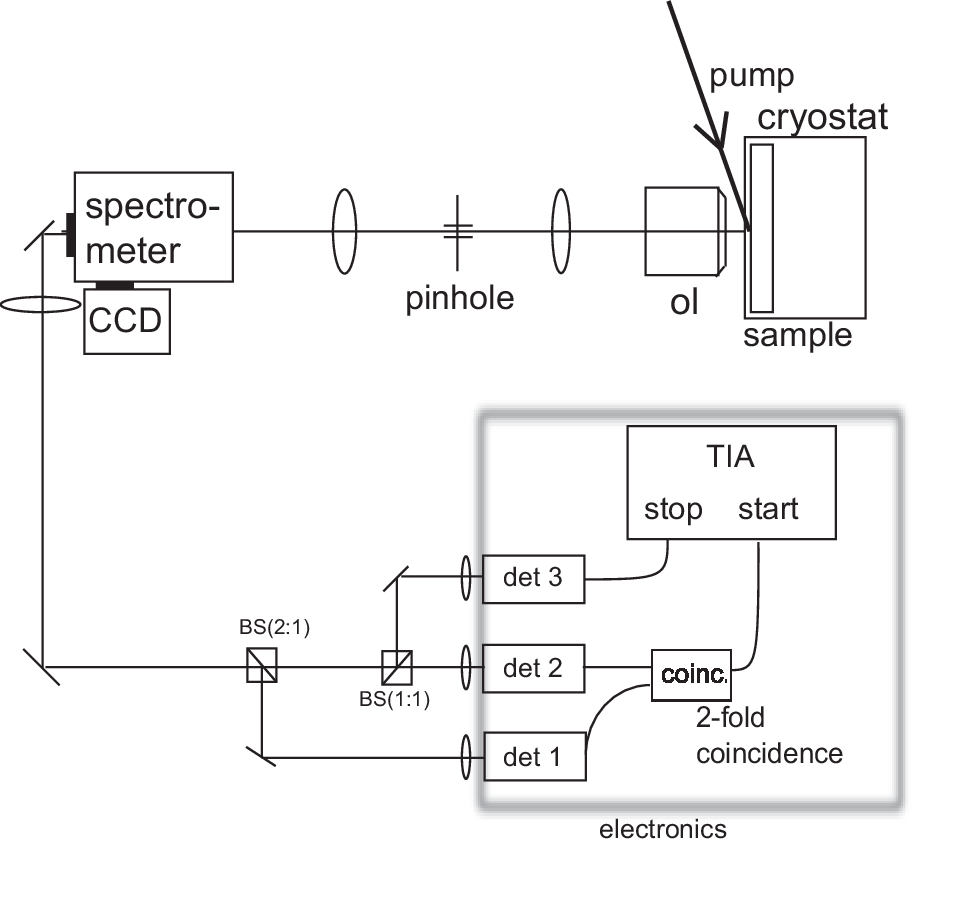}}}
    \caption{Experimental setup. pump: Mode-locked Ti:Sa laser with 76 
    MHz repetition rate and 4 ps pulse width. GaAs QW microcavity 
    sample is hold in a cryostat which keeps the sample temperature at 10 
    K during experiment. ol is an objective lens with 4 mm focal 
    length and 0.55 NA. Momentum distribution is imaged at the entrance 
    slit of the spectrometer by the ol and the following two lenses. 
    Thus, dispersion curve of LP can be directly observed by the attached 
    CCD camera and LP ground state emission is spectrally selected.
    When the signal goes through the other exit, it enters HBT setup. 
    First two beam splitters (BS(2:1) and BS(1:1)) splits the signal into 
    three paths with roughly equal intensities for maximizing 3-fold 
    coincidence.
    Electronics consists of electrical 
    processing units for taking 2-fold or 3-fold coincidences, i.e. 
    single photon detectors (det 1 to det 3), delay units for timing 
    adjustments,
    discriminator (delay units and discriminator are not shown in the 
    figure), coincidence
    unit, and time-interval analyzer (TIA).
    }\label{1}
   \end{center}
  \end{figure}

 In the present work, we used 12 GaAs QWs embedded in GaAlAs/AlAs distributed Bragg 
 reflector (DBR) microcavity that was need in our previous experiments 
 \cite{chiwei, utsunomiya}.
  The sample has three 
  stacks of four GaAs QWs which are embedded at the central three 
  antinode positions of a DBR 
  planar microcavity. Lateral trapping potential is provided by a hole 
  surrounded by a thin metal (Ti/Au) film which pushes 
  photon-field amplitude antinode position at GaAlAs/air interface 
  inside GaAlAs layer, resulting in the blue shift of the cavity resonance 
  and hence LP energy under the metallic layer. 
  By using this metallic hole structure, we could obtain the confinement 
  of polaritons under a hole, which allows us to access to a single
  spatial mode in second and third order coherence measurements.
  An optical measurement system together with electronics for third 
  order coherence 
  measurement is shown in Fig. \ref{1}. 
  A mode-locked Ti:sa 
  laser with 4 ps pulse width is utilized as a pump laser. It is focused on 
  the sample surface with diameter of about 50 $\mu$m. An objective lens 
  collimates the output luminescence and  near field image is made at the 
  position of the pinhole by the second lens. Here we can spatially filter 
  the metallic hole area of about 5 $\mu$m diameter. Then at the entrance slit of the spectrometer, 
  far field is imaged by the third lens.
  Dispersion relation can be measured by 
  the spectrometer and an attached nitrogen-cooled CCD camera 

\begin{figure}[Hh]
   \begin{center}
    \rotatebox{0}{
    \scalebox{0.4}{
    \includegraphics{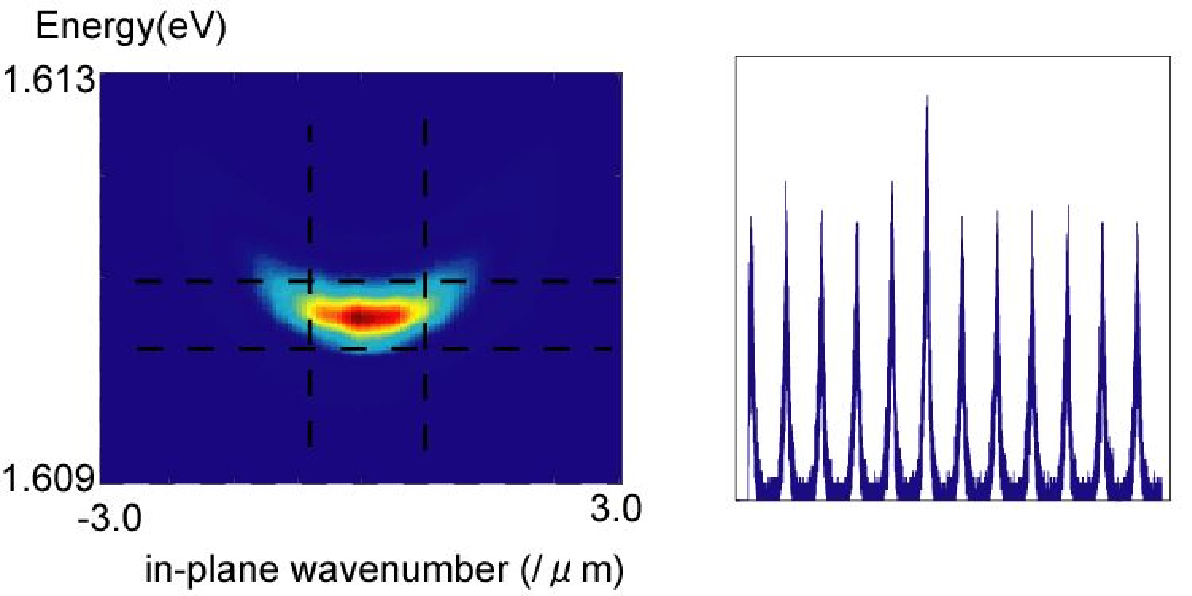}}}
    \caption{(left) Energy vs wavenumber dispersion of LP around 
    $4P_{\rm th}$. Closed area 
    surrounded by dotted lines is picked up by using spectrometer and 
    appropriate collection optics. (right) An example of 3 fold coincidence TIA data. Note that 
    though it looks the same as usual 
    $g^{(2)}(0)$ measurement, this histogram is triggered by input signals 
    from 2-fold coincidence unit, therefore, this histogram gives information 
    of $g^{(3)}(0)/g^{(2)}(0) $. The highest peak 
    corresponds to detections of three photons from one pump pulse. 
    Surrounding peaks are 2-fold coincidence by one pump (start signal 
    of TIA) but the stop 
    signal of TIA is generated by other pump pulses at different time slots.
    Interval between each peak is 13 ns corresponding to the repetition 
    rate of pump laser. 
    }\label{2}
   \end{center}
  \end{figure}

  The left-hand side of Figure \ref{2} is 
  an example of observed dispersion at $4P_{\rm th}$ ($P_{\rm 
  th}$ is threshold power ). 
  In case  we use the side exit of the monochromator, 
  the output signal enters Hanbury-Brown Twiss (HBT) setup consisting of 
  two non-polarizing beam splitters and photon detectors. We can measure 
  $g^{(2)}(0)$ and $g^{(3)}(0)$ with this setup.
@Signal entering HBT 
  setup is split into three paths by a 2:1 splitting ratio
  beam splitter and a 1:1 ratio beam splitter. We used single photon 
  counting modules (Perkin Elmer, SPCM AQR series) for this coherence 
  function measurement.
  The detectors' response time is relatively slow (about 300 ps 
  timing resolution), but its slow response does not limit our 
  capability of measuring $g^{(2)}(0)$ and $g^{(3)}(0)$
  since we  used a pulsed pump laser of 4 ps pulse width and the 
  emission time becomes shorter than the correlation time at well above threshold. 
  This setup allows us to pick up arbitrary part of condensates at LP ground 
  state by setting appropriate wavelength filter of spectrometer.
  The in-plane momentum region is determined by collecting 
  optics and actual CCD image size.
  Area surrounded by dotted line in Figure \ref{2} is the detected 
  region. 

  For $g^{(3)}(0)$ measurement, we need to take 3-fold coincidence of three
  detectors. For normalization of coherence function,
  3-fold coincidence rate of simultaneous three photon
  detection for one pump pulse must be divided by accidental 
  signal (product of independent signal rates of three detectors).\
  In the actual measurement, we used a
  time interval analyzer (TIA) which measures time difference between start 
  and stop inputs. $g^{(2)}(0)$ can be directly measured by TIA but 
  $g^{(3)}(0)$ cannot be since our TIA has just 2 input channels (start and stop). 
  $g^{(3)}(0)$ is instead measured by the following method. At first, we took 
  2-fold coincidence between detector 1 and detector 2 by 2-fold 
  coincidence unit. Then the output signal is input to start port of the 
  TIA (Fig. \ref{1}). Right-hand side of Figure \ref{2} is an example of raw 
  histogram taken by TIA. The horizontal axis corresponds to delay 
  time between start signal and stop signal. The interval between 
  next pulses is 13 ns which corresponds to repetition period of the 
  pump laser.
  The highest peak corresponds to 3-fold coincidence of simultaneously 
  detected three photons for one pump pulse and surrounding peaks 
  correspond to 2-fold coincidence by simultaneously detected two photons 
  for one pump pulse and accidental third photon from different time slot.
  So the surrounding peaks 
  correspond to $g^{(2)}(0)$, i.e. just detector 1 and detector 2 are fired 
  by the same pulse. In this configuration, we need to 
  preliminary know $g^{(2)}(0)$ to obtain $g^{(3)}(0)$ since taking the ratio 
  between central peak and average of surrounding peaks just gives 
  $g^{(3)}(0)/ g^{(2)}(0)$. So $g^{(3)}(0)$ can be obtained by 
  multiplying $g^{(2)}(0)$ to the ratio of 
  central peak to surrounding peak.
  
\begin{figure}[Hh]
   \begin{center}
    \rotatebox{0}{
    \scalebox{0.3}{
    \includegraphics{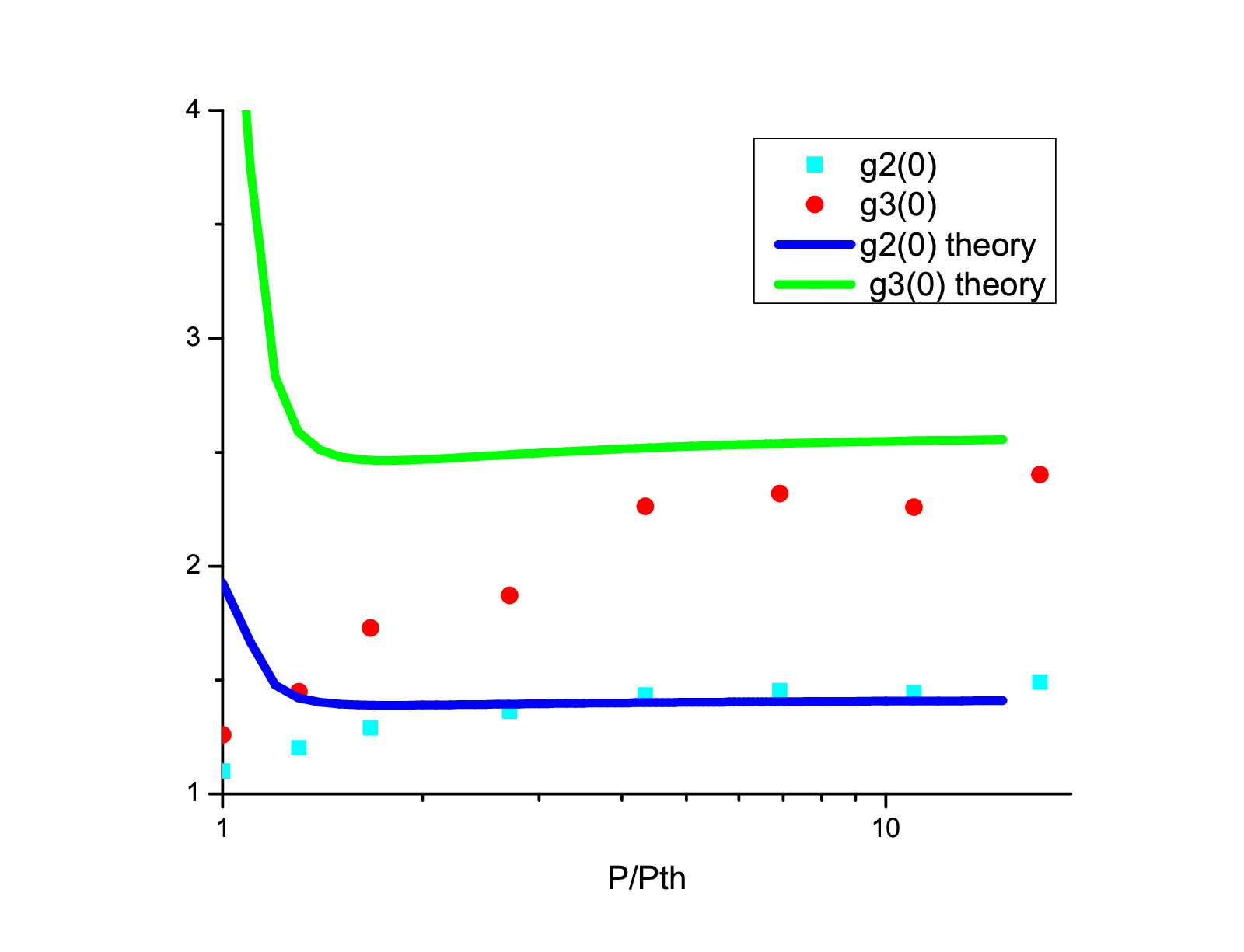}}}
    \caption{Pump intensity dependence of $g^{(2)}(0)$ and $g^{(3)}(0)$. $P_{\rm 
    th}$ is threshold power. square:$g^{(2)}(0)$, and circle:$g3(0)$. Two 
    curves are calculated coherence functions whose theoretical model is 
    described in \cite{sch2}.
    }\label{result}
   \end{center}
  \end{figure}

  We measured $g^{(2)}(0)$ and $g^{(3)}(0)$ at various pump powers
  (Fig. \ref{result}).
  In Fig \ref{result}, theoretical results are drawn together with 
  experimental data. The details of the theoretical model are given in Ref. \cite{sch2}.
  Below threshold, though they are not shown 
  in Fig \ref{result}, the statistics obeys thermal distribution, hence, 
  the theory predicts $g^{(n)}(0)= n!$. Just above threshold, they begin to gain coherence and 
  rapidly decrease towards unity. However, the decreases stop around 
  $1.5 P_{\rm th}$ due to increasing effect of polariton scattering. 
  As the pump intensity increases, they converge into certain values 
  ($g^{(3)}(0)\sim2.5$ and $g^{(2)}(0)\sim1.4$).
  The experimental data
  shows $g^{(2)}(0)$ and $g^{(3)}(0)$ are still close to unity just above 
  threshold.
  However, they increase and become closer to theoretical values 
  as the pump intensity increases. After gradual increase, they reach 
  a flat area at around $4P_{\rm th}$. 

  To understand the discrepancies between theory and 
  experiment at the pumping regime of $1<P/P_{\rm th}<4$, 
  we need to consider intensity correlation time and decay time of the 
  condensate emission.

\begin{figure}[Hh]
        \begin{center}
    \rotatebox{0}{
    \scalebox{0.25}{
    \includegraphics{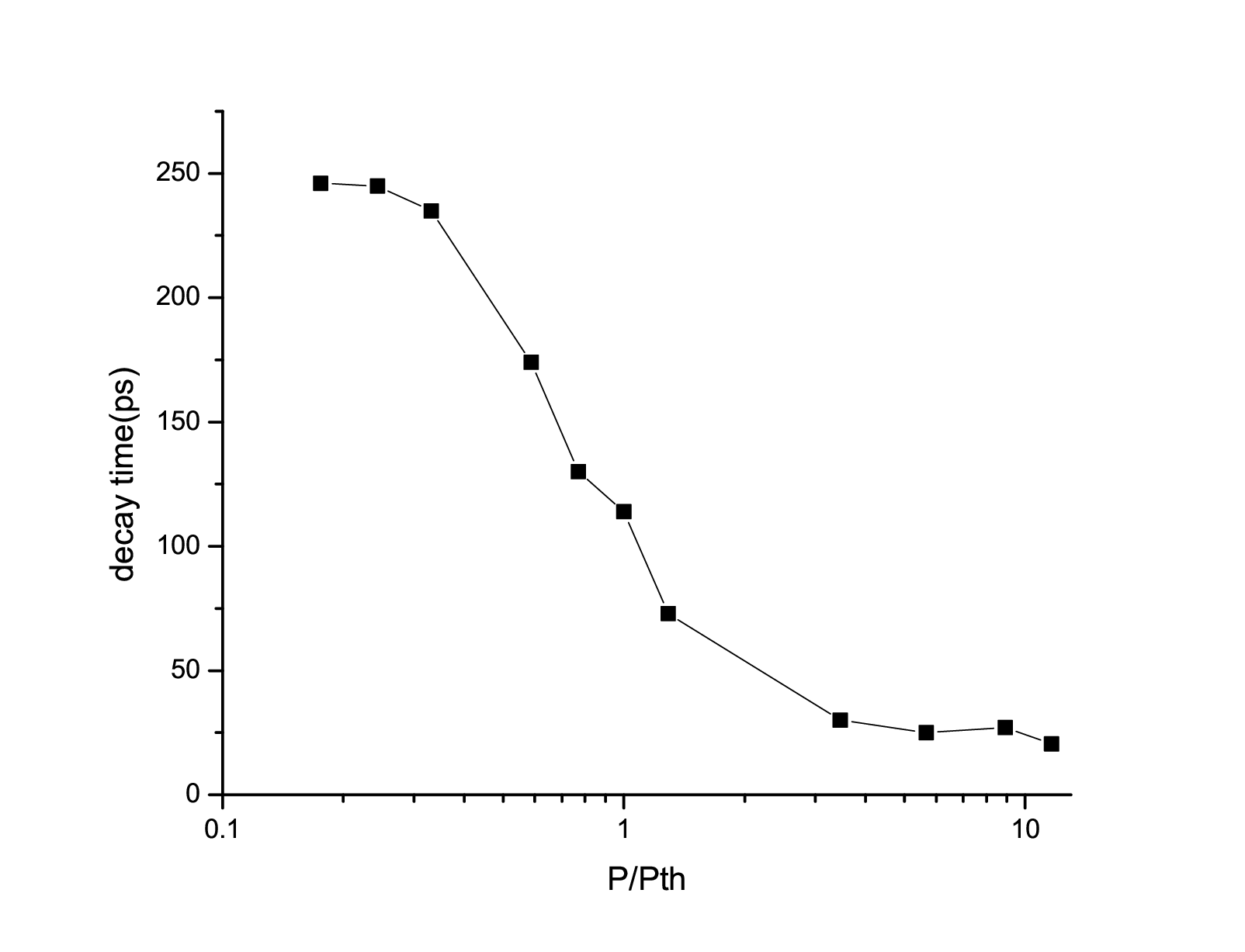}}}
    \caption{A photoluminescence decay time of a LP ground state. Around 
	 threshold, the decay time gradually decreases. It reaches a
	 minimum value of 20 ps around 3-4 $P_{\rm 
	 th}$ where $g^{(2)}(0)$ and $g^{(3)}(0)$ 
	 approach to the theoretical values. }
	 \label{decay}
	 \end{center}
  \end{figure}

  The ground state 
  photoluminescence decay time measured by a streak camera preceded by a 
  spectrometer is shown in Fig. \ref{decay}. Since the streak camera was set 
  after a spectrometer, we could avoid the contamination by the 
  non-condensates and the wavelength range giving maximum PL intensities 
  at each pump power was picked up.
  Below threshold where photon statistics is expected to obey thermal 
  distribution, the observed statistics is always 
  $g^{(2)}(0)= g^{(3)}(0) \approx1$ since decay time $\tau $ of photoluminescence is much 
  longer than the intensity correlation time $\Delta \tau_c$. In the case of longer 
  $\tau \gg \Delta \tau_c$, $ g^{(n)}(\tau)=1$ \cite{hui}. The measured 
  $g^{(n)}$ is the 
  integration of $g^{(n)}(\tau)$ over an integration time window which 
  is given by an emission time $\tau$ in our case,
  , hence, it is close to unity after averaged over the whole 
  emission lifetime. Just above threshold, the condition is still the same since 
  the emission pulse width is longer than the intensity
  correlation time until the pump rate reaches $4P_{\rm th}$. Finally 
  far above 
  threshold, the intensity correlation time becomes closer to the pulse width, 
  and then the intrinsic noise property of the condensates begins to be detected properly. 
  

  In conclusion, we have experimentally measured the second and third order 
  coherence functions. The observed bunching effect
  is the experimental evidence for the relatively strong 
  thermal and quantum depletion of the LP
  condensate. This higher order coherence 
  function measurement technique may contribute to a 
  further investigation of coherence property of an
  exciton-polariton condensate.

T.H. thanks G. Roumpos, N. Y. Kim, and S. Utsunomiya for their help. And 
T.H. and Y. Y. acknowledge financial support from NICT, Special 
Coordination Funds for Promoting Science and Technology, MEXT and DARPA.

\end{document}